# Anomalous normal state resistivity in superconducting $La_{2-x}Ce_xCuO_4$: Fermi liquid or strange metal?

Tarapada Sarkar[1], Richard L. Greene[1] and S. Das Sarma[2]

[1]*Center for Nanophysics & Advanced Materials and Department of Physics, University of Maryland, College Park, Maryland 20742, USA.*
[2]*Condensed Matter Theory Center and Joint Quantum Institute, Department of Physics, University of Maryland, College Park, Maryland 20742, USA.*

**Abstract**

We present experimental results for the in-plane resistivity of the electron-doped cuprate superconductor $La_{2-x}Ce_xCuO_4$ above its transition temperature $T_c$ as a function of Ce doping $x$ and temperature. For the doping $x$ between 0.11 and 0.17, where $T_c$ varies from 30 K ($x$=0.11) to 5 K ($x$=0.17), we find that the resistivity shows an approximate $T^2$ behavior for all values of doping over the measurement range from 100 K to 400 K. The coefficient of the $T^2$ resistivity term decreases with increasing $x$ following the trend in $T_c$. We analyze our data theoretically and posit that n-type cuprates are better thought of as strange metals. Although the quadratic temperature dependence appears to be in naive agreement with the Fermi liquid (FL) expectations, the fact that the measured resistivity is large and no phonon-induced linear-in-$T$ resistivity manifests itself even at 400 K argue against a standard normal metal Fermi liquid picture being applicable. We discuss possible origins of the strange metal behavior.

Ever since the discovery of the cuprate superconductors with high superconducting transition temperatures 1986 [1], one of its most studied properties has been the dc transport in the normal phase for $T>T_C$. It is universally acknowledged that the temperature dependent resistivity, $\rho(T)$, of cuprates is not understood since that there is no consensus on the underlying scattering mechanism responsible for the temperature dependent resistivity. The terminology 'strange' metal has been invoked specifically to describe the 'metallic' temperature dependence of cuprates. This strange metallicity, which is not uniquely defined, , refers to a linear-in-$T$ resistivity manifesting over a large temperature range as well as the absolute high-temperature magnitude of the resistivity often exceeding the so-called Mott-Ioffe-Regel (MIR) limit ($\sim 150$ $\mu\Omega$-cm in usual metals) [2-5] without showing any sign of resistivity saturation. Of course, $\rho(T) \sim T$ at room temperatures is nothing special as all normal metals behave this way, but the high absolute value of $\rho(T)$ in cuprates (approaching 1 m$\Omega$-cm at 600K or so) [4,5] is a serious conundrum. The fact that the linearity of $\rho(T)$ sometimes persists to rather low temperatures has often been emphasized as being strange as well, but this in fact often happens in normal metals too-- e.g. $\rho(T)$ in Cu is linear from $T=50$K to 600K [6]. In fact, in high-mobility two-dimensional semiconductors, a linear-in-$T$ metallic resistivity may persist down to 1K or below because of the low Fermi momentum [7, 8]. It is generally believed that strange metallicity arises from strong correlations, and perhaps, from a hidden quantum critical point under the superconducting dome in the cuprates, but there is no widely accepted theory for strange metals, and certainly none of the various proposed ideas leads to a calculated linear-in-$T$ (and very large) resistivity in agreement with the experimental data in any real materials. Strange metallicity has been proposed to be a manifestation of non-FL behavior associated with the non-existence of quasiparticles although no agreed upon calculation, starting from a microscopic model of the cuprates, has been able to quantitatively explain the observed transport properties of the strange metal. Understanding the characteristic properties of strange metals, particularly in the context of cuprates, remains an open and important challenge in theoretical physics (see, e.g., Ref. [9] for an up to date discussion of this issue).

In the current work, we present experimental data on the normal state temperature-dependent resistivity of electron-doped LCCO (La$_{2-x}$Ce$_x$CuO$_4$) in the $T$= 30-300 K range (for $x$=0.11 to 0.17) showing that in this particular cuprate system (with a maximum $T_c$ of ~30 K at x=0.11), $\rho(T)$ obeys a clear $T^2$-law up to 400 K. Below 80 K, the power law exponent drops rapidly from 2 approaching 1 around $T$~ 30 K just above $T_c$, but the existence of superconductivity makes the analysis of the temperature dependence of $\rho(T)$ problematic for $T$=30-80 K at zero magnetic field. Our measured absolute values of the resistivity are very high (in the 200-500 μΩ-cm range at 300 K) and being > 100 times larger than the room temperature resistivity of simple metals, LCCO does indeed qualify for the designation of a strange metal in spite of its quadratic temperature dependence. In fact, we posit that our observed quadratic $T$-dependence is actually stranger than the ubiquitous linear $T$-behavior since all elemental metals manifest a linear-in-$T$ temperature dependence in $\rho(T)$ in this temperature range, and none shows a quadratic $T$-dependence in the 80-400K temperature range as we find for LCCO. In contrast to hole-doped cuprates [10], we find there is no doping (x) within the SC dome where conventional metallic FL behavior can explain our zero-field normal state data. Our new results are in contrast with previous studies on n-type cuprates which claimed the $\rho \sim T^2$ behavior above ~ 100 K consistent with FL theory [11,12]. These earlier conclusions were based on a limited range of doping compared to the comprehensive data that we present here.

Our main experimental resistivity results are shown as four (a-d) panels in Fig. 1. The measured in-plane resistivity (fig. 1a) varies from ~50 μΩ-cm (x=0.11) to ~10 μΩ-cm ($x$=0.17) at low temperatures (just above $T_C(x)$) and from ~450 μΩ-cm ($x$=0.11) to ~200 μΩ-cm ($x$=0.17) at 300K. Writing the in-plane resistivity as $\rho = \rho_0 + \Delta\rho(T)$, we subtract out the extrapolated $T$=0 residual resistivity $\rho_0$, and the T-dependent contribution to the in-plane resistivity is found to vary as $\Delta\rho \sim T^2$ for $T$=80-300K regime (see ref [38] fig. S1c). The logarithmic derivative in Fig. 1b is rather definitive in showing a plateau in the effective temperature exponent to be close to 2 in the whole 100-300 K range and for $x$=0.11-0.17. The direct fittings are shown in the supplementary (see Ref. [38] (Fig. S1). We mention that the exponent value (~2) does not change even if the full resistivity $\rho(T)$ is fitted to the $T^2$ behavior without the subtraction of the residual resistivity. The residual resistivity $\rho_0$ and the SC transition temperature $T_c$ are shown as a function of the doping $x$ in Fig. 1d, where a clear trend of $\rho_0$ and $T_c$ both decreasing with increasing $x$ can be seen. One aspect of strange metallicity is indeed this dichotomy of increasing resistivity correlating approximately with increasing $T_c$ on the optimal to overdoped regime as seen in Fig. 1d. We note that the residual resistivity $\rho_0$ is indeed much smaller (by a factor of 5-10) than the temperature dependent part $\Delta\rho(T)$, making the extraction of the exponent 2 meaningful. Based on ref -13 and the measured data for two samples up to 400 K as shown in Fig. 2, the roughly quadratic (n=1.85±0.02) temperature dependence of the LCCO resistivity likely persists in the whole 100-800K temperature range in the optimal to overdoped regime. Somewhere at higher $T$ (~ 800K and above), $\rho(T)$ likely crosses over to a linear-in-$T$ behavior [13], but this is beyond the scope of the current work. The 'transport' phase diagrams for our system ($x$=0.11-0.17) based directly on our results (Fig. 1 and Fig.2) and including data from the existing literature. is shown in Fig. S2 [38].

In Fig. 3 we analyze the T-dependent resistivity shown in Fig. 1c by writing $\rho(T) = \rho_0(x) + A(x)T^2$, and plotting $A(x)$ as a function of $x$. (The dependence of $\rho_0$ on $x$ is shown in Fig. 1d.) It is obvious from Fig. 3 that $A(x)$ decreases with increasing x, following an approximate power law, $A \sim 1/x^\alpha$, with the exponent α ~2.6. We note that $T_C(x)$ (also shown in Fig. 3) decreases with decreasing $A(x)$, i.e., increasing $x$. The behavior $A(x) \sim x^{-\alpha}$ with α ~ 2.6 may have significance with respect to the resistive scattering mechanism leading to the observed quadratic increase of the in-plane resistivity with temperature as discussed below.

The key question arising from our data is the nature of the underlying scattering mechanism causing the quadratic temperature dependence in the in-plane resistivity, which is uncommon in cuprates [10,11,13-18], and has sometimes been reported in other strongly correlated materials [19-21]. First, we note that in simple elemental metals (e.g. Cu, Al), ρ(T) in the 100 K-300 K range is invariably linear, arising entirely from the electron-acoustic phonon interaction. This is in fact the expected generic behavior of resistivity in any electronic Fermi liquid material in the presence of acoustic phonons at 'high' temperatures, $T > T_D/5$, where $T_D$ is the lattice Debye temperature. Logically, there are three distinct ways of escaping this linear-in-$T$ phonon-induced metallic resistivity: (1) the system is a strange metal, and consequently, a non-Fermi-liquid where the excitations do not couple to phonons in the usual metallic manner; (2) the characteristic phonon temperature $T_D$ is very high ($T_D$> 3000 K) so that phonon modes are frozen out and contribute little to ρ(T) in the 100 K- 300 K range; (3) some other scattering mechanism dominates phonon scattering in the 100 K- 300 K regime leading to a $T^2$ law in the resistivity. Which one of these three reasons is operational in LCCO is unknown, but it is reasonable to rule out (2) and ignore (1) for discussion thus focusing on item 3. The Debye temperature in LCCO is likely to be ~ 400 K-500 K [22], and all discussions will stop if we accept item (1) since non-Fermi-liquids are a clever way of saying that we do not understand at all what is going on.

The third possibility has a natural candidate because of the $T^2$ temperature dependence, which immediately suggests electron-electron interactions as the mechanism for scattering. This is what has been claimed in all prior resistivity studies of n-type cuprates [23]. But we will argue below that this is unlikely to be the correct interpretation. The FL theory for electron scattering provides the scattering rate going as $1/\tau \sim T^2 f(n)$, where $f(n)$ is a function of the effective carrier density of the system. In the leading order theory, $f(n) \sim n^{-\gamma}$, where $\gamma \sim 1$ in the leading order perturbation theory [24]. The net resistivity of the system, assuming it to be a Fermi liquid, is then given by: $= \frac{m}{n\tau e^2}$, where 'm' is the carrier effective mass. Assuming the carrier effective mass 'm' to be independent of the doping $x$ and the effective carrier density 'n' to be proportional to doping $x$, we then conclude that $\rho(T) \sim A(x)T^2$, where $A(x) \sim x^{-2}$. It is interesting to note that this highly simplified theory gives a dependence of $\rho(T)$ which agrees with the $T^2$ dependence found in Fig. 1, and also gives reasonable agreement with $\alpha \sim 2.6$ found in Fig. 3. The Fermi liquid value for $\alpha = \gamma + 1 \sim 2$, is different from the experimentally obtained exponent $\alpha \sim 2.6$, but this difference is rather small given the highly simplified nature of the theory. For example, one expects some dependence of the effective mass on $x$ and the simple n ~ $x$ dependence may not be quantitatively valid, possibly leading to the 25-30% difference between $\alpha$ and $\gamma + 1$. This scenario is also consistent with the fact that $\rho_0(x)$ itself in Fig. 1(d) falls off somewhat faster than 1/x indicating that the carrier density n does not follow the simple x~n linear relationship since the residual resistivity, being dependent only on quenched disorder, should vary as 1/n. Taking into account a stronger than linear dependence of n on x, the value of

α~2.6 is consistent with the electron-electron interaction prediction of 1+ γ=2. Another possibility is that there is an additional $log(\frac{T}{T_F})$ [25-27] term in 1/τ for two dimensional systems, which could, in principle, modify the exponent bringing theory and experiment in closer agreement. In addition, the large measured resistivity can be understood simply from the low carrier density of LCCO (expected to be around $10^{21}$ per cm$^3$) [28], which leads to a rather small value of the Fermi temperature $T_F$~ 2000 K, leading to a rather small τ arising from electron-electron interaction (and hence a large resistivity). In fact, assuming n~ $10^{21}$ per cm$^3$ (and m given by the free electron mass) gives a $\rho(T)$ within a factor of 2 of the measured value, which is remarkable given the simplified nature of the theory and the approximate values of the system parameters. We add that the effective scattering time here under the same assumptions is comparable to the scattering time at 300 K in normal metals (~ $10^{-14}$ s), and the high resistivity (by a factor of 100 compared with normal metals) may be arising primarily from the effective low carrier density (~ $10^{21}$ cm$^{-3}$) in the system. One sign of strangeness in our system is the observed $T^2$ dependence of the scattering rate for values of $T/T_F$ ~ 0.3-0.5 since the low carrier density (and high effective mass) here implies a rather low value of $T_F$ —in a FL, the interaction induced $T^2$ dependence of carrier scattering manifests only for $T/T_F \ll 1$.

The semiquantitative, i.e. α~γ+1 agreement between the quadratic temperature dependent (~$T^2$) experimental resistivity and the FL theory assuming electron-electron interaction (~$T^2$) to be the underlying scattering mechanism seems to have clinched the matter in favor of FL physics leading to transport in LCCO. However, this is a very incomplete picture since electron-electron interaction, being momentum conserving, should not directly affect the resistivity unless a momentum conservation breaking mechanism acts in concert. Such mechanisms could be umklapp scattering or Baber scattering [29], but there is no particular reason to believe that such processes play important roles in LCCO, compared, e.g., with hole doped cuprates where $\rho(T)$ is often linear [9,30]. In fact, in normal simple metals, the $T^2$ Fermi liquid resistivity has never been cleanly observed experimentally, and therefore, it may be a bit presumptuous to attribute our observed $T^2$ dependence to FL electron scattering physics. A full theory of electron-electron scattering in LCCO including umklapp processes is well beyond the scope of the current work. We also note that conventional FL physics implies that the optical conductivity ($\sigma_1(\omega)$) should have an $\omega^2$ dependence in the temperature range where the dc resistivity is proportional to $T^2$. This is not the case for $\sigma_1(\omega)$ in optimally doped $Pr_{1.85}Ce_{0.15}CuO_4$ [31] where the $T^2$ dc resistivity is also observed between 100-400 K [13] as shown in figure 2b.

    We do mention one possibility for electron-electron scattering to affect the resistivity in our system. It is hydrodynamics, i.e., strong inter-electron collision happening at a much faster rate than either electron-impurity or electron-phonon scattering so that the system is in a local equilibrium [32-34]. It is possible in such a hydrodynamic fluid for electron-electron scattering induced $T^2$ resistivity to manifest itself in the electrical resistivity, and we believe that, if electron scattering is indeed the underlying mechanism responsible for producing the quadratic temperature dependence, then hydrodynamics may be a more reasonable scenario than umklapp or Baber scattering in our system. Considering a full hydrodynamic theory for LCCO transport is a formidable challenge well beyond the ability of current theories, but we should mention the fact that our observed $\Delta\rho(T) \gg \rho_0$ and the absence of any phonon-induced linear-in-$T$ resistivity in the data are consistent with the quadratic temperature dependent arising from hydrodynamic effects since clearly electron-impurity and electron-phonon scatterings are weak in our system. This hypothesis is strongly supported by the recent observation of an unconventional (i.e non-FL) thermal diffusivity between 200-600 K in optimally doped NCCO

crystals [35], the same temperature range where $\rho \sim T^2$ is observed [13]. Hydrodynamics may also explain the $T^2$ dependence persisting to high T values consistent with our observation in contrast to a standard FL.

It may be useful to comment on any possible role of quantum criticality here since the possible existence of a hidden quantum critical point under the SC dome near optimal doping has been a well-discussed theoretical theme in the cuprate physics literature, including the behavior of the normal state resistivity. In LCCO, there is most likely a quantum phase transition involving a FSR around x~0.14[28], which may be relevant for the observed quadratic temperature dependence since the precise effect of such a critical point on the resistivity is unknown except that it is widely believed to produce power laws in temperature. Why such a hidden critical point would produce a $T^2$ power law in LCCO versus a $T$ power law in hole-doped cuprates [9] is unknown, and beyond the scope of the current work. We mention, however, that our experimental temperature exponent is ~1.85 (see Ref [38] for doping x=0.11 to 0.17), and therefore the possibility that there is a small component of a linear-in-$T$ resistivity contribution in addition to the dominant $T^2$ term cannot be ruled out. The fact that the exponent decreases (most likely) to around unity for T>800K could be arising from either phonon contributions (as in normal metals) or from the hidden quantum criticality (as in other cuprates), which cannot be discerned without further work. Another possibility for a crossover to an effective linear $T$ behavior at very high temperatures ($T$>800K) could be that the resistivity saturation effect, which invariably manifests itself as a suppression of $d\rho/dT$ with increasing $T$, is operational as the resistivity approaches the putative MIR limit ~ 2 m.ohm.cm for the low-density LCCO system at higher temperatures.

Finally, we comment on the implication of our data for strange metallicity. In general, strange metallicity and Fermi liquid behavior are thought to be mutually exclusive, and strange metals (Fermi liquids) are often defined by linear (quadratic)- in-$T$ temperature dependence of the electrical resistivity along with very large resistivity. We believe that this dichotomy is unfounded since simple metals, the quintessential Fermi liquids (e.g. Cu, Al, Ag), manifest linear-in-$T$ resistivity, and almost never a quadratic-in-$T$ resistivity. Thus, operationally the observation of a quadratic $T$-dependence in the resistivity for $T$=100-400K (and possibly up to 800K) is much more strange than the observation of a linear-in-$T$ resistivity in the same temperature range since the linear behavior is routine in every simple metal (and obviously, one cannot argue that simple metals are non-FL). The really challenging question is why in LCCO (and also $Nd_{2-x}Ce_xCuO_4$ (NCCO) [12,13], $Pr_{2-x}Ce_xCuO_4$ (PCCO) [13] and hole doped cuprates in the pseudogap regime [10,30]), the resistivity manifests a $T^2$ resistivity at room temperatures whereas the standard FL (i.e. simple metals) manifest only a linear-in-$T$ resistivity. In addition, as argued above, electron interactions should not affect the resistivity of a simple Fermi liquid unless an underlying momentum conservation breaking mechanism is operational. If hydrodynamics is indeed what is causing the quadratic $T$-dependence, then our observations are extremely interesting because a strongly interacting hydrodynamic quantum electron fluid is indeed a strange metal although it is not a non-FL in the sense of not having a Fermi surface at zero temperature [32]. Also, a related hydrodynamics must cause a linear T-dependence in the hole doped cuprates! Obviously, much more work will be necessary for settling this important question, but we have sharpened the theoretical question here considerably: Is LCCO a simple FL or is it a strongly interacting hydrodynamic quantum fluid? [36,37] Or is the quadratic temperature dependence arising from something completely different (e.g. quantum criticality associated with a FSR) whose origin is unknown at this stage. The results and the analysis presented here argue strongly against a simple FL picture.

**Acknowledgement:** We thank N. Butch, J. Paglione and Aharon Kapitulnik for discussions. This work is supported by the NSF under Grant No.DMR-1708334, the Maryland "Center for Nanophysics and Advanced Materials and the Laboratory for Physical Sciences (SDS).

**Figure Captions:**

**Figure 1:** Resistivity for La$_{2-x}$Ce$_x$CuO$_4$ (LCCO) thin films with $x$=0.11,0.13,0.14,0.15,0.16 and 0.17. (a) *ab*-plane resistivity versus temperature for all x; (b) $\frac{dlog(\Delta\rho(T)=\rho(T)-\rho(0))}{dlog(T)}$ vs Temperature; (c) *ab*-plane resistivity versus quadratic temperature (color) with $\rho(T) = \rho(0) + AT^n$ fit (solid black line) from 40 K to 200 K; (d) $\rho(o)$ (extracted by extrapolating to $T=0$ the normal state resistivity by a polynomial fit) vs doping (black) and $T_c$ vs doping ( blue). Red, blue and black error bars are the uncertainty of the doping, $\rho(o)$ and $T_c$ respectively.

**Figure 2:** (a) Resistivity for x=0.15(black), 0.17(blue) LCCO films measured up to 400 K fitted with $\rho(T) = \rho(0) + AT^n$ (red) where n=1.80±0.02. b) ab-plane resistivity versus quadratic temperature for Nd$_{1.85}$Ce$_{0.15}$CuO$_4$ (blue square) and Pr$_{1.85}$Ce$_{0.15}$CuO$_4$ (black square) with $\Delta\rho(T) = AT^2$ fit (red solid line) (data taken from ref-13)

**Figure 3:** Magnitude of $T^2$ resistivity, $A(x)$ taken from fig 1(c) vs doping (black) fitted with $A(x) \alpha\, x^{-\alpha}$ (magenta) and $T_c$ vs doping (blue) with $\alpha = 2.61 \pm 0.054$. Red error bar is uncertainty in the doping and black error bar is uncertainty in $T_c$

# Figures

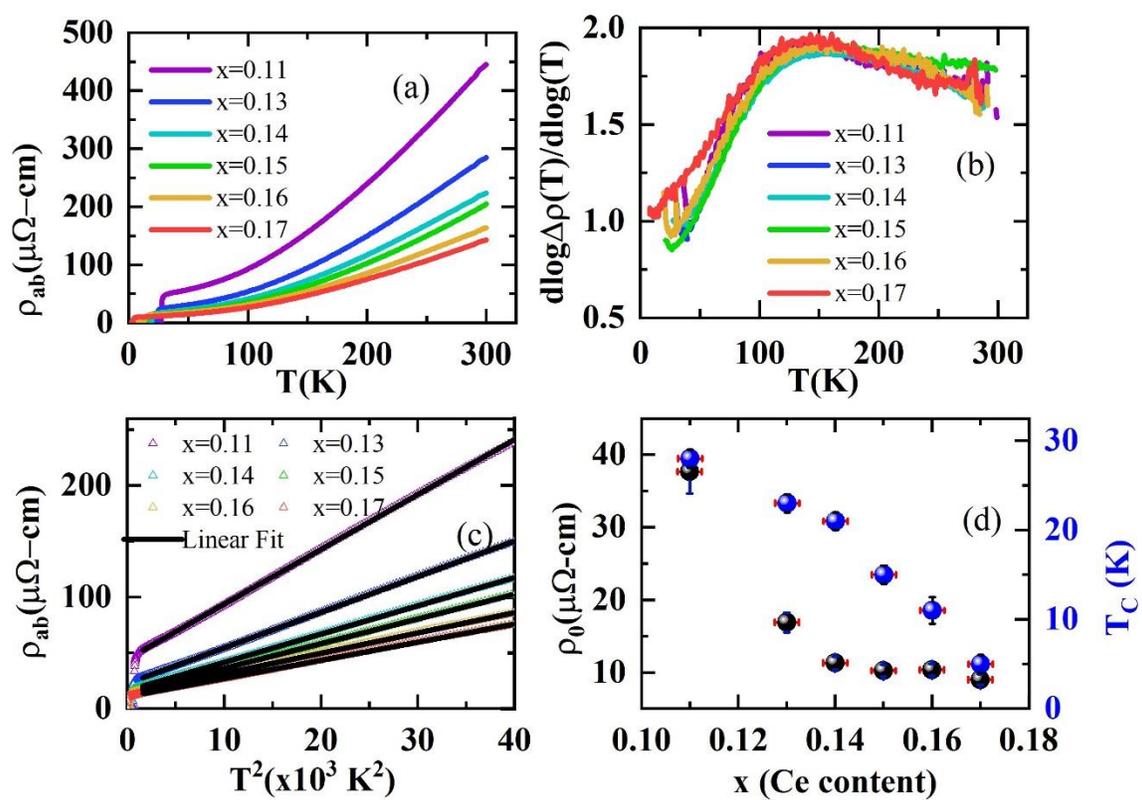

**Figure 1. Tarapada Sarkar et al.**

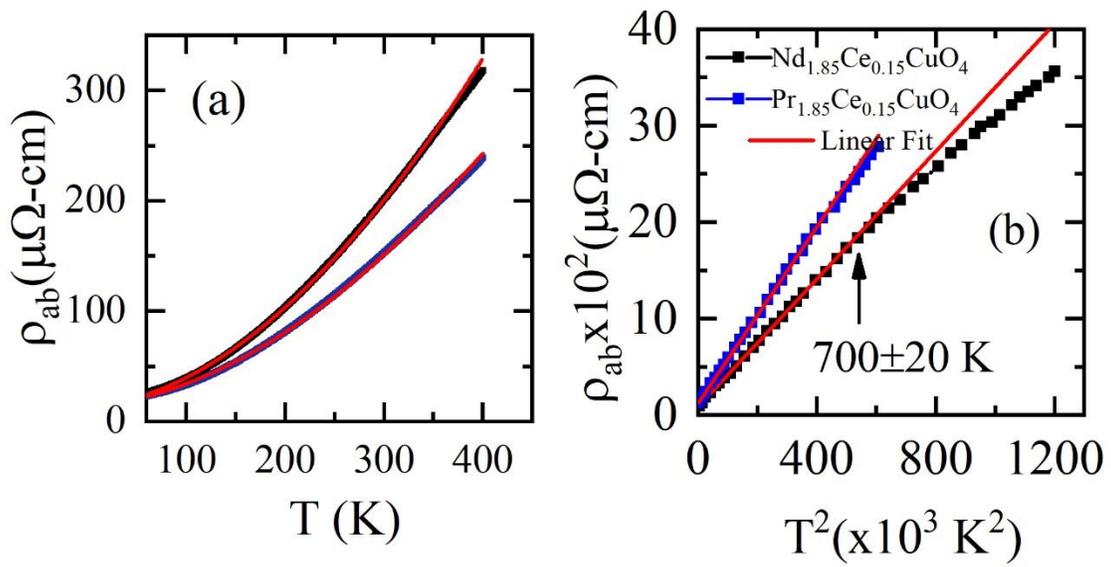

**Figure 2. Tarapada Sarkar et al.**

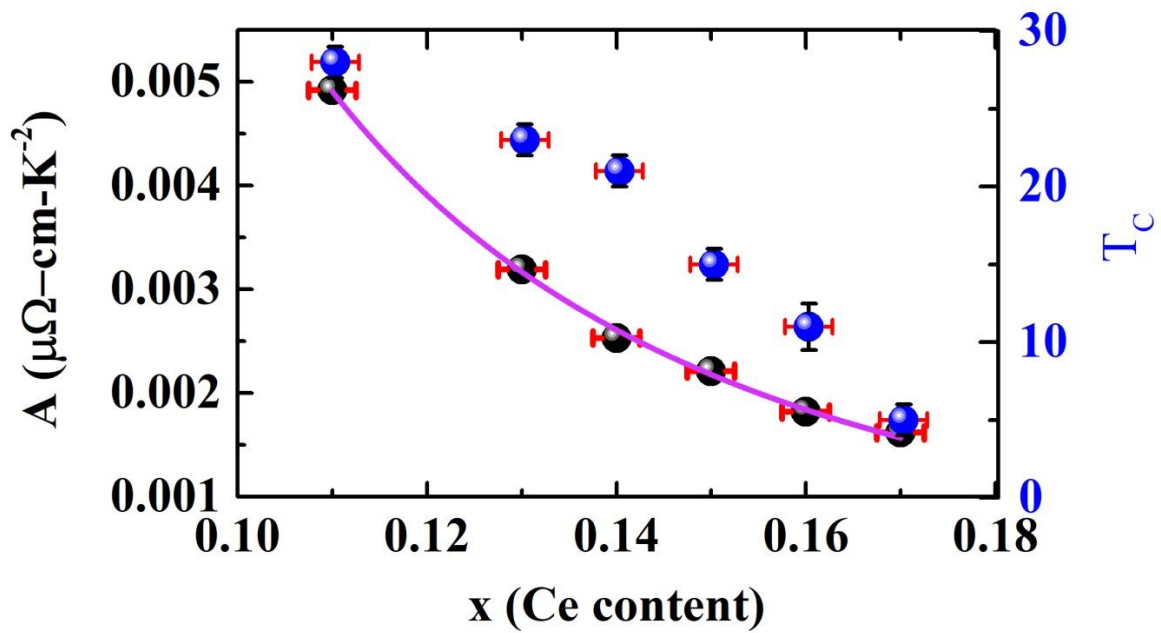

**Figure 3. Tarapada Sarkar et al.**

## Supplemental Material

**Methods**

The measurements have been performed on $La_{2-x}Ce_xCuO_4$ films for optimally doped ($x$=0.11) and overdoped ($x$=0.13, 0.14, 0.15, 0.16, 0.17) compositions. High quality LCCO films (thickness about 150 - 200 nm) were grown using the pulsed laser deposition (PLD) technique on $SrTiO_3$ [100] substrates (5×5 mm$^2$) at a temperature of 700 °C utilizing a KrF excimer laser. The full width at half maximum of the peak in $d\rho_{xx}/dT$ of the films is within the range of 0.2-0.8 for the optimum and overdoped films, demonstrating the high quality of the samples. The LCCO targets have been prepared by the solid-state reaction method using 99.999% pure $La_2O_5$, $CeO_5$, and CuO powders. The Bruker X-ray diffraction (XRD) of the films shows the c-axis oriented epitaxial LCCO tetragonal phase. The thickness of the films has been determined by using cross sectional scanning electron microscopy (SEM). The resistivity measurements of the lithographically pattern films have been performed in a Quantum Design Physical Property Measurement System (PPMS).

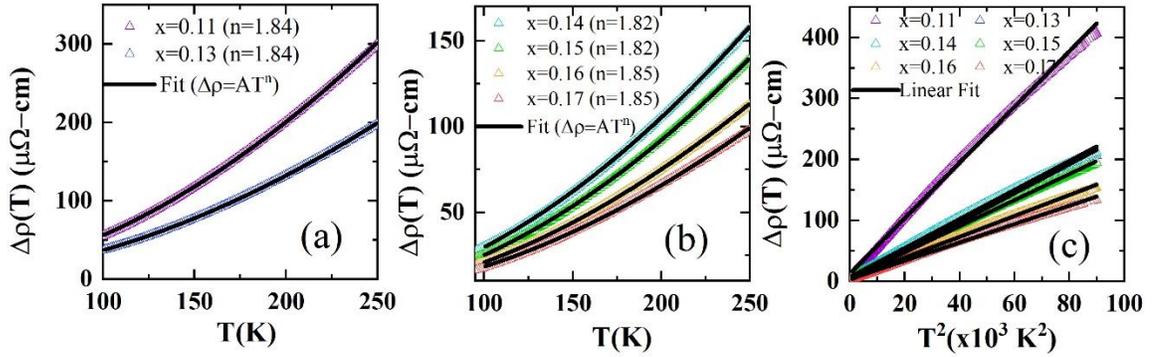

**Figure** S1(a) and (b) $\Delta\rho(T)$ vs $T$ for all x (color) from 100 K to 250 K with $\Delta\rho(T)=AT^n$ fitting (solid black line); (c) $\Delta\rho(T)$ vs $T^2$ from $T_C$ to 300 K for all x (color) with $\Delta\rho(T) = AT^2$ fit (solid black line) from $T_C$ to 300 K.

In figure S1 we show the quadratic fits (in Figs. S1(a,b)) and direct fitting of the resistivity data to the $T^2$ law (linear fits in Figs. S1(c) which shows excellent agreement in Fig. 2b). Although our current LCCO samples and the measurement set-up preclude us from going to $T$ >400 K, we speculate, based on Fig.2, that the quadratic temperature dependence of the LCCO resistivity likely persists in the whole 100-800K temperature range in the optimal to overdoped regime. Somewhere at higher $T$ (~ 800K and above), $\rho(T)$ likely crosses over to a linear-in-$T$ behavior (as is apparent in Fig. S2), which, however, is beyond the scope of the current work.

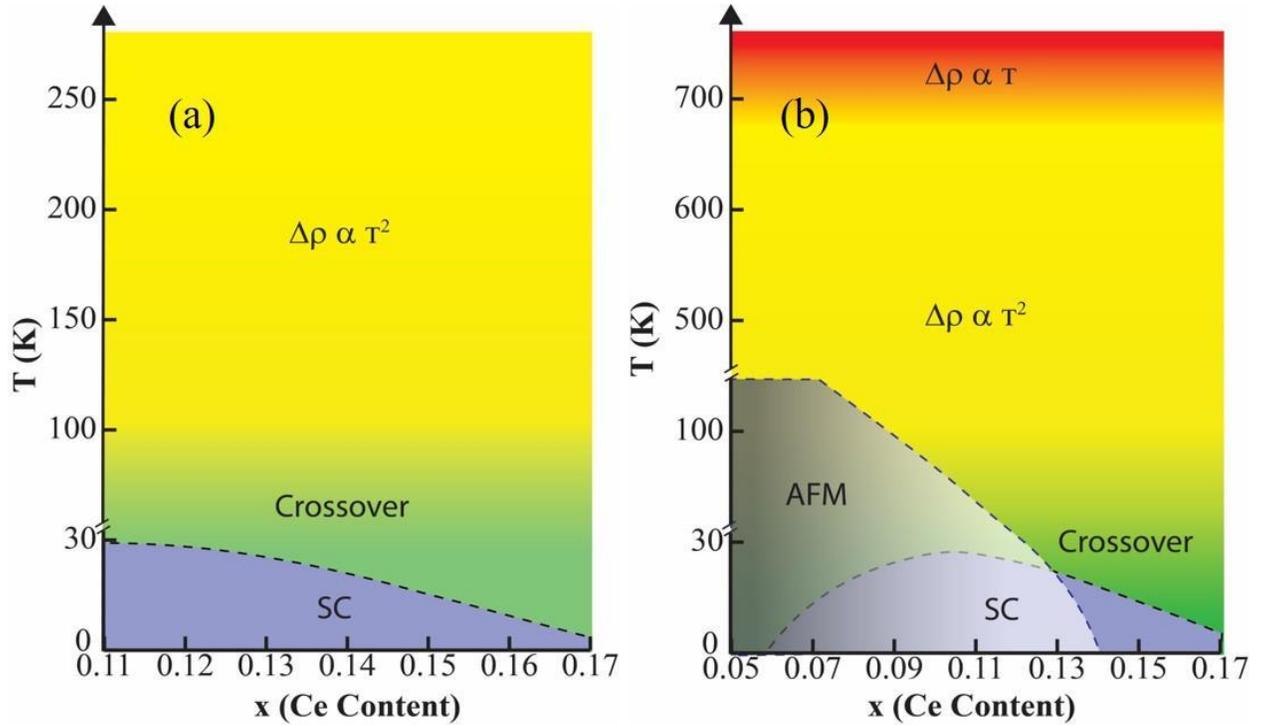

**Figure S2:** Schematic temperature–doping phase diagram of LCCO. (a) Phase diagram in zero magnetic field ($H = 0$) from this work. The superconducting phase (light blue) lies below $T_c$ (dotted black line), normal state (yellow regime) starts from a crossover (light green) above 100 K. (b) Schematic full phase diagram of LCCO. Long range and short range AFM order (ref-2) ends at x=0.14 (ref-28 main text), quadratic temperature dependent resistivity (yellow regime) ends around 700 K (as suggested by ref-1), where the linear temperature dependent resistivity (red regime) crossover starts.